\documentstyle[12pt]{article}
\begin{document}
\input psfig
\def \mpla#1#2#3{Mod. Phys. Lett. A {\bf#1}, #2 (#3)}
\def \nc#1#2#3{Nuovo Cim. {\bf#1}, #2 (#3)}
\def \np#1#2#3{Nucl. Phys. {\bf#1}, #2 (#3)}
\def \pisma#1#2#3#4{Pis'ma Zh. Eksp. Teor. Fiz. {\bf#1}, #2 (#3) [JETP Lett.
{\bf#1}, #4 (#3)]}
\def \pl#1#2#3{Phys. Lett. {\bf#1}, #2 (#3)}
\def \plb#1#2#3{Phys. Lett. B {\bf#1}, #2 (#3)}
\def \pr#1#2#3{Phys. Rev. {\bf#1}, #2 (#3)}
\def \prd#1#2#3{Phys. Rev. D {\bf#1}, #2 (#3)}
\def \prl#1#2#3{Phys. Rev. Lett. {\bf#1}, #2 (#3)}
\def \prp#1#2#3{Phys. Rep. {\bf#1}, #2 (#3)}
\def \ptp#1#2#3{Prog. Theor. Phys. {\bf#1}, #2 (#3)}
\def \rmp#1#2#3{Rev. Mod. Phys. {\bf#1}, #2 (#3)}
\def \nim#1#2#3{Nucl. Instru. \& Meth.{\bf #1}, #2 (#3)}
\def \rp#1{~~~~~\ldots\ldots{\rm rp~}{#1}~~~~~}
\def \yaf#1#2#3#4{Yad. Fiz. {\bf#1}, #2 (#3) [Sov. J. Nucl. Phys. {\bf #1},
#4 (#3)]}
\def \zhetf#1#2#3#4#5#6{Zh. Eksp. Teor. Fiz. {\bf #1}, #2 (#3) [Sov. Phys.
-JETP {\bf #4}, #5 (#6)]}
\def \zpc#1#2#3{Zeit. Phys. C {\bf#1}, #2 (#3)}
\def\etal{et al.}

\begin{titlepage}
\hspace*{\fill}HEPSY 95-2\linebreak
\hspace*{\fill}Sept., 1995~~~\linebreak

\large
\centerline {\bf CLEO III, A Dectector To Measure Rare $B$ Decays}
\centerline{and CP Violation}
\normalsize

\vskip 2.0cm
\centerline {Sheldon Stone\footnote{Stone@suhep.phy.syr.edu}}
\centerline {\it Physics Dept., Syracuse Univ., Syracuse, NY, 13244-1130}
\vskip 4.0cm

\centerline {\bf Abstract}\vskip 1.0cm
The symmetric $e^+e^-$ collider CESR is undergoing a series of upgrades
allowing for luminosities in excess of $2\times 10^{33}$cm$^{-2}$s$^{-1}$.
The most important goals of the upgrade are precision measurement of standard
model parameters $V_{cb}$, $V_{ub}$, $V_{td}/V_{ts}$, $f_{D_s}$, and
searching for CP
violation and standard model violations in rare $B$ decays. A new detector
upgrade, called CLEO III, has started which includes a new silicon-wire
drift chamber tracking system and a Ring Imaging Cherenkov Detector, RICH,
using a LiF radiator and CH$_4$-TEA gas based photon detector.

\vspace{1cm}
\begin{flushleft}
.\dotfill .
\end{flushleft}
\begin{center}
{Presented at BEAUTY '95 - 3nd International Workshop on B-Physics at Hadron
Machines Oxford, UK, 10-14 July, 1995}
\end{center}
\end{titlepage}

\newpage
\section{Introduction - $B$ Physics Goals}

 The CLEO collaboration is in the midst of a detector upgrade,
to match the CESR increase in luminosity to $> 2\times
10^{33}$cm$^{-2}$s$^{-1}$
 and to insure meeting the physics goals described below.

\begin{figure}[hbt]
\vspace{-.8cm}
\centerline{\psfig{figure=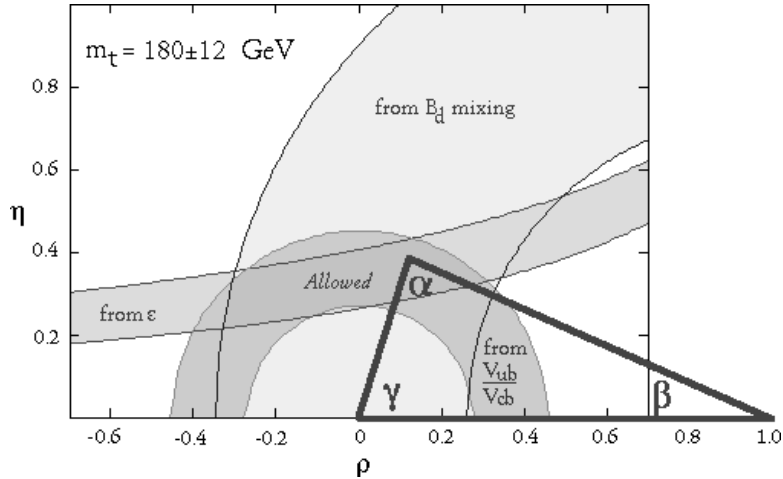,height=3in,bbllx=0bp,%
bblly=425bp,bburx=600bp,bbury=800bp,clip=}}
\vspace{-1.0cm}
\caption{\label{CKMtri}The CKM triangle overlaid upon constraints in the
$\rho-\eta$
plane, from measured values of $V_{ub}/V_{cb}$, $B^0-\bar{B^0}$ mixing
and $\epsilon$ in the $K^0$ system. The allowed region is given by the
intersection of the three bands.}
\end{figure}

The current extent of knowledge on weak mixing in the quark sector can
be shown by plotting constraints in the
 $\eta$ and $\rho$ plane given by measurements of the $\epsilon$ parameter
describing CP violation in $K^0_L$ decay,
and by $B^0-\bar{B}^0$ mixing and semileptonic $b\to X\ell\nu$ decays. An
analysis of the allowed parameter space is shown in
Figure~\ref{CKMtri} \cite{Pascos}.
Overlaid on the figure is a triangle that results from the requirement
$V\cdot V^{\dag} = 1$, i.e. that the CKM matrix be unitary. Measurements of CP
violation in $B$ decays can, in principle, determine each of the angles
$\alpha$, $\beta$ and $\gamma$ of this triangle independently.

There are many measurements to make with CLEO III. Here I will only
mention a few of the most important.
\begin{itemize}
\item The dominant error in the $\epsilon$ band is caused by the uncertainty
in $V_{cb}$. This can be improved, for example, by measuring the rate for
$B^-\to D^{*0}
\ell^-\nu$ at maximum four-momentum transfer.
\item The error in the $V_{ub}$ band can be decreased by using exclusive
semileptonic decays such as $\pi\ell^-\nu$ \cite{pilnu}.
\item The error on the mixing band comes dominately from the lack of
knowledge of $f_B$. We can reduce this uncertainty by comparing models
with our measurement of $f_{D_s}$ currently ($284\pm 30\pm 30\pm 16$) MeV
\cite{fDs}.
\item The ratio $V_{td}/V_{ts}$ may be ascertained by measuring the ratio
of the $\rho\gamma/K^*\gamma$ branching ratios \cite{rhogam}. The basic
idea here is that the difference in the two final
states comes mainly from having $V_{ts}$ in the case of $K^*\gamma$
and $V_{td}$ for $\rho\gamma$. Although it has
been argued that other diagrams may contribute to the numerator this would
lead to the inequality of $\rho^+\gamma$ and $\rho^-\gamma$ and would
therefore be a measurement of direct CP violation. It is also likely that
these other diagrams would induce an inequality between neutral and charged
$\rho$ final states. In short, this is a very interesting physics area.
\item Measure CP violation. We will try and measure the CP violation angle
$\gamma$ using one of
two techniques. The first, suggested by
 Gronau and London, involves measuring the rates for
$B^-\to D^0K^-$, $\overline{D}^0K^-$ and $D_{CP}^0K^-$, and the corresponding
rates for $B^+$ decay. ($D_{CP}^0$ indicates that the $D^0$ decays into a
CP eigenstate, for example $K^+K^-$.) The second involves measuring a rates
for two-body pseudoscalar decays $\pi^{\pm}\pi^0$ and $K^{\pm}\pi^0$
etc...\cite{rhogam}.
Other attempts to measure CP violation through rate asymmetries such as
differences between $\psi\pi^+$ and $\psi\pi^-$ or even to measure the angle
$\beta$ by using the standard $\psi K_s$ mode and the small $B$ motion in
our symmetric $e^+e^-$ collider.
\end{itemize}

\section{CESR upgrades}

 CESR has achieved the highest instantaneous luminosity of any
$e^+e^-$ collider, $\approx 2\times 10^{32}$cm$^{-2}$s$^{-1}$. However,
heavy quark physics is limited by the number of events, so even higher
luminosities are desired. The approved CESR upgrade will have luminosity
in the range of 2$\times 10^{33}$. This is accomplished in two stages. In the
first stage there will be 9 ``trains" of 3 bunches, and a small horizontal
crossing angle, with a goal of $6\times 10^{32}$. This phase is being
implemented now. The second stage has 9 trains of 5 bunches, a larger
horizontal crossing angle, a smaller $\beta_v^*$ ($\approx$1 cm) requiring a
new
interaction region and therefore a new inner detector, and superconducting
RF cavities.

There is also a future plan being developed, to reach luminosity in excess of
$5\times 10^{33}$ by using 180 bunches, round beams in the interaction
region and a reduced bunch length. This plan would use the same
inner detector being built now, which is the subject of the rest of this
article.

\section{The CLEO III Dectector Upgrade}

The CLEO II goal was to reconstruct thousands of $B$ mesons using excellent
tracking and photon detection and modest particle identification.
Although the CLEO II goals have been met, experience with the detector and the
desire to  meet new physics goals, have uncovered several serious deficiencies.
These include the rather thick drift chamber endplate and associated material
which seriously compromises photon detection in the endcaps and outer barrel
region, and the lack of good quality charged hadron identification\footnote{The
Drift Chamber dE/dx system does provide $\sim 2\sigma$
$K/\pi$ separation above 2.3 GeV/c.}above 750
MeV/c. Furthermore, the electronics and DAQ
system will not support the higher interaction rates caused by the increased
luminosity. Finally, the inner tracking detector must be replaced because
 part of the space is
necessary for the final focus quadrupoles.

There are several advantages to the  detector being upgraded since CESR is
a symmetric energy accelerator. It is easier to have a large solid angle
coverage than in an asymmetric machine, since less of the phase space goes
down the beam pipe. The
maximum momentum of a $B$ decay product is 2.8 GeV/c, rather than about
4 GeV/c for an asymmetric machine operating at the $\Upsilon(4S)$. It is
also important that we can build upon our existing detector utilizing some
of the more expensive parts such as the CsI calorimeter, the magnet and the
muon system, all of which have functioned excellently.

The design philosphy of CLEO I was to have excellent tracking and charged
particle momentum resolution. Advances in crystal growth techniques,
superconducting magnet
technology and electronics made it possible to add excellent photon and
electron detection in CLEO II \cite{CLEOII}. The design philosophy of CLEO III
is to
maintain the excellent charged particle resolution while sacrificing 20
cm of space for charged particle identification. This is possible because
we can add a high precision silicon detector near the beam pipe which can
precisely determine angles and then have a low mass drift chamber which
can precisely measure transverse momentum. It is important to realize that
multiple scattering is the largest source of tracking errors over the entire
range of $B$ decay momenta at CESR.

\begin {figure} [htbp]
\vspace{-1.cm}
\centerline{\psfig{figure=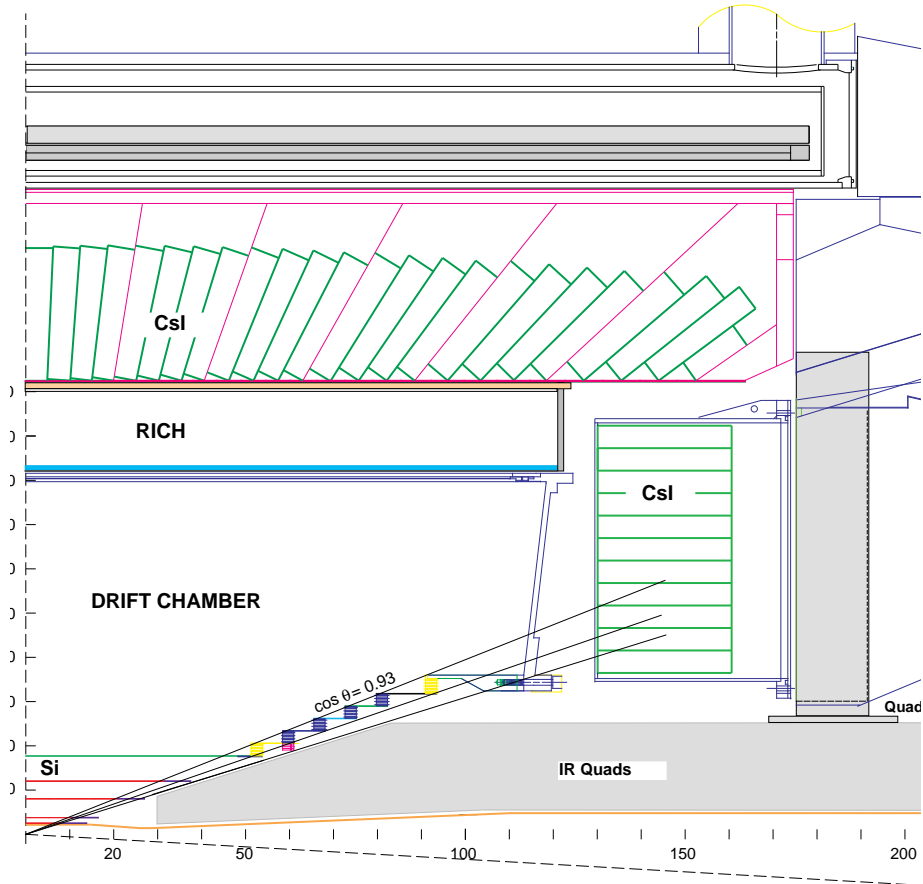,height=4.5in,%
bbllx=37bp,bblly=40bp,bburx=611bp,bbury=558bp}}
\vspace{-2.5cm}
\caption{\label{CLEOIII}Quarter section of the CLEO~III detector.}
\end {figure}

An overview of CLEO III is shown in Fig.~\ref{CLEOIII}.
  There is a four layer silicon detector,
followed by a drift chamber with staggered inner section to accommodate
the CESR final focus quadrupoles.  Radially beyond this is 20 cm space
for the barrel particle identification detector.  At the ends of the
detector are relocated and repackaged CsI calorimeter endcaps with improved
performance and provision for mounting
the drift chamber and CESR machine components.

\section{The RICH system}
\subsection{Introduction}

Ring imaging Cherenkov detectors (RICH) are capable of providing
excellent identification of charged particles.
Several systems have been implemented in hadron beams and $e^+e^-$ collider
experiments \cite{overv}.

\begin{figure}[hbt]
\vspace{-1cm}
\centerline{\psfig{figure=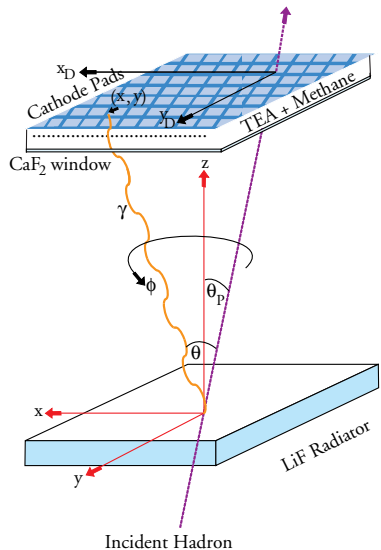,height=4in,bbllx=0bp,%
bblly=0bp,bburx=600bp,bbury=700bp,clip=}}
\vspace{-2.8cm}
\caption{\label{RICHscm} Schematic diagram of LiF-TEA RICH system.}
\end{figure}

The CLEO RICH is based on the
`proximity focusing' approach, in  which the Cherenkov cone is simply let to
expand in a volume filled with ultraviolet transparent gas
(the expansion gap) as much as allowed by  other spatial constraints, before
intersecting the detector surface where the  coordinates of the Cherenkov
photons are reconstructed.
The components of our
system are illustrated in Fig.~\ref{RICHscm}. It consists of a LiF radiator,
providing U.V.
photons, an expansion region, and a photosensitive detector made from
a wire proportional chamber containing a CH$_4$-TEA gas mixture. The cathode
plane is segmented into pads to reconstruct the photon positions.

 The design is based on work of the
College de France--Strasbourg group,\cite{CERNt}.
We are considering a unique geometry for the LiF solid radiator \cite{alex},
where the light emitting surface is shaped like the edge of a saw blade
with the teeth angled at 45$^o$. If we can manufacture this shape it will
produce more photons and achieve better resolution.
A fine  segmentation of the cathode pad ($\approx$ 7.5x8mm$^2$) is required in
order to achieve
the spatial  resolution needed, which in turn implies a high density of
readout electronics.

The angular resolution per detected photon is comprised of several sources.
The most important are the chromatic error,  which results from the variation
of the index of refraction with the wavelength, the emission point error,
which results from the lack of knowledge about where the photon is
emitted, and the position error in detecting the photon. There is
also a small error caused by the inability to resolve some of the
overlapping photon cluster.
The individual sources of error have been determined by using GEANT. The
combined resolution is
about 13.5-14 mrad resolution per detected photon independent of
the track incident angle. This corresponds to a 3.7 mrad resolution per track.
 Since the difference in Cherenkov angle between pions and kaons is 12.8 mr at
2.8 GeV/c we expect this system to have at least 3.5 standard deviation
 separation. However, should it be possible to obtain the sawtooth radiators,
 the  separation will be much better \cite{alex}.

The device will be constructed with a 30 fold segmentation.  The LiF radiators
are held in place by an inner carbon fiber cylinder to which they are
attached.
   The back of the cathode boards which constitute
the outer side of the MWPC are   strengthened by hollow G10 rods which
also act as channels for the  cooling  gas ($N_2$).
The strength has been achieved with great care to minimize the amount of
material in the detector, in order to preserve the excellent performance of
the electromagnetic CsI calorimeter. The average material thickness is
13\% of a radiation length for tracks at normal incidence.

\subsection{ Readout Electronics}

 The 230,000 readout channels are distributed over the
surface at the outer radius of the
detector and are impossible to access routinely. Therefore the
readout architecture must feature high parallelism, and
extensive testing of active components and
connection elements is required in order to insure their reliability.
 The MWPC detector surface is segmented into 30 modules, which will be
divided into 12 subunits each with 640 readout channels. Each of
these subunits will communicate via a low mass cable connection with VME
cards providing the control signals and receiving the analog or
digitized signals as discussed below.

Several considerations affect the design of the individual channel
processor. Low noise is an essential feature because the charge
probability distribution for the avalanche produced by a single photon
is exponential at moderate gains. It should be stressed that it
is beneficial to run at low gains to improve the stability
of chamber operation. Therefore in order to achieve
high efficiency, the noise threshold should be as low as
possible. An equivalent noise charge of about 200 electrons is adequate
for our purposes. On the other hand, an exponential distribution implies
 that a high dynamic range is desirable in order to preserve the
spatial resolution allowed by charge weighting.
Note that charged  particles are expected to  generate pulses at least 20 times
higher than the single photon mean pulse height.
In order to improve the robustness of the readout electronics against
sparking, a protection circuit constituted by a series resistor and two reverse
biased diodes is required in
 the input stage. Finally it is
 important to sparsify the information as soon as possible in the
processing chain, as the occupancy of this detector is very low and
therefore only a small fraction of the readout channels contain useful signals.

There is a preamplifier/shaper VLSI chip developed for
solid state detector applications which incorporates many of the
features discussed above [2]. A dedicated version of this chip,
called VA\_RICH, has been developed and will be tested shortly. Its predicted
equivalent noise
charge is given by:
\begin{equation}
ENC=\sqrt{(73 e^- + 12.1*C e^-/pF)^2+(50e^-)^2},
\end{equation}
where $C$ is the input capacitance of $\approx$2 pf.
The first term corresponds to the noise contribution from the
input transistor and the (80 $\Omega$) series resistor used for the input
protection
 and the second  to the noise from
subsequent stages, small but non negligible because of the lower gain
chosen to increase the dynamic range. This device is expected to maintain
linear response up to an input charge of 700,000 $e^-$.

The choice of digitization and sparsification technique has not yet been
finalized. Under consideration is the digitization and sparsification at the
front end level, using the zero suppression scheme and the ADC included
in the SVX II readout chip \cite{chip}. Alternatively we will digitize all the
analog output signal and perform the zero suppression afterwards.

\subsection {Results from Prototype}

We have constructed a prototype of an individual detector module
about 1/3 the length of an actual detector module and about the same width.
 The prototype system is enclosed in a leak tight
aluminum box. The
expansion gap is 15.7 cm. In this prototype we have used plane 1 cm thick
LiF radiators.
 The chamber geometry is approximately the same as the final design
in terms of gap size, wire to cathodes distance and pad sizes. The total
number of pads read out is 2016. Pad signals are processed
by VA2 preamplifier and shapers \cite{einar}. The detector plane is divided
into 4 quadrants each of which has 8 VA2 daisy--chained for serial readout.

 \begin{figure}[hbt]
\vspace{-1cm}
\centerline{\psfig{figure=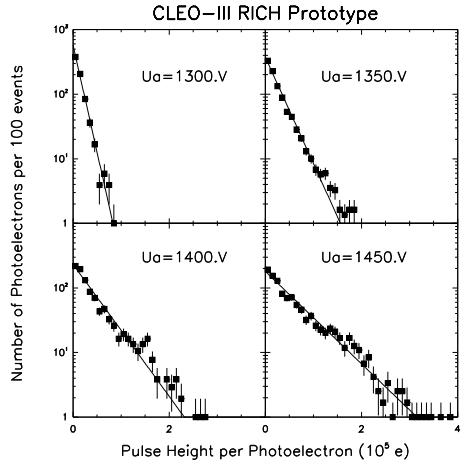,height=3.5in,%
bbllx=0bp,bblly=0bp,bburx=600bp,bbury=700bp,clip=}}
\vspace{-2cm}
\caption{\label{phdis} Photon induced avalanche charge distribution at
different anode
voltages. The voltage on the metallization of the CaF$_2$ windows is
kept at -1350V.}
\end{figure}

\begin{figure}[hbt]
\vspace{-1.cm}
\centerline{\psfig{figure=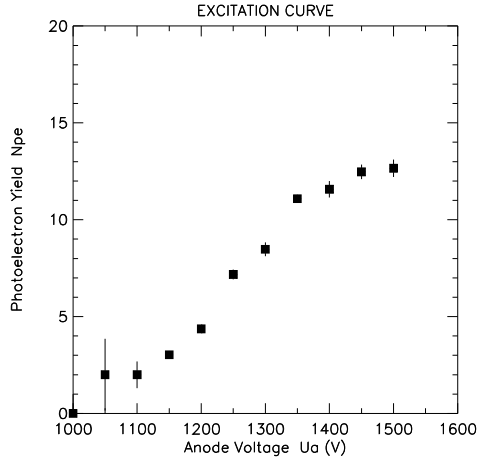,height=3.5in,bbllx=0bp,%
bblly=0bp,bburx=600bp,bbury=700bp,clip=}}
\vspace{-2cm}
\caption{\label{plateau} Excitation curve for CH$_4$-TEA. The voltage on the
metallization
of the CaF$_2$ windows is kept at -1350V.}
\end{figure}

This prototype has been exposed to hardened cosmic rays.
Fig.~\ref{phdis} shows the charge distribution of reconstructed photon clusters
as
a function of the anode wire voltage $U_a$ for several different voltages.
Note that the pulse height distribution is consistent with an exponential
profile and its mean value increases with $U_a$, as expected.
Fig.~\ref{plateau}
shows the excitation curve for hits which are more than six standard
deviations above noise. It can be
seen that the plateau corresponds to $N_{pe}\approx 13$. This number has
to be corrected for possible background hits, which we estimate to be
about 1 per event.
 This performance is in close agreement with
our expectations based on the performance of a similar prototype built
and tested by the College de France--Strasbourg group \cite{CERNt}. A
tracking system is being added to this set--up to allow us to start measuring
Cherenkov angular resolutions.

\section{Silicon vertex detector/tracker}

In order to provided good measurements of angles and precise vertex
positions, useful especially for charm decays, we are constructing
a four layer double sided silicon device, which covers 93\% of 4$\pi$.
Our concept is to use only one detector size with 50 $\mu$m strip pitch
and external dimension of 2.7cm$\times$5.26cm$\times$300$\mu$m.

The number of detectors and their properties are listed in
Table~\ref{table:si}.
The large amount of capacitance on the fourth layer makes it difficult to
achieve low noise performance. The noise target is 125$e^-$ + 8$e^-$/pf in
a radiation hard version of the chip. The rad-hard version of the Viking chip,
VH1, gives 400$e^-$ + 5$e^-$/pf, which is almost good enough. The silicon
group is pursuing making their own front end preamplifier.

\begin{table}[th]
\centering
\caption{CLEO III silicon detector properties}
\label{table:si}
\vspace*{2mm}
\begin{tabular}{rrrrrc}\hline\hline
%\multicolumn{2}{c}{$\BM$}&\multicolumn{2}{|c}{$\BP$}
Layer\# & r(cm) & \# in $\phi$ & \# in z & total \# & C(pf)\\\hline
1 & 2.4 & 7 & 4 & 28 & 12\\
2 & 3.8 & 10 & 4 & 40 & 19\\
3 & 7.5 & 20 & 8 & 160 & 38 \\
4 & 12.0 & 30 & 12 & 360 & 60 \\
\end{tabular}
\end{table}

\begin{figure}[hbtp]
\vspace{-2cm}
\centerline{\psfig{figure=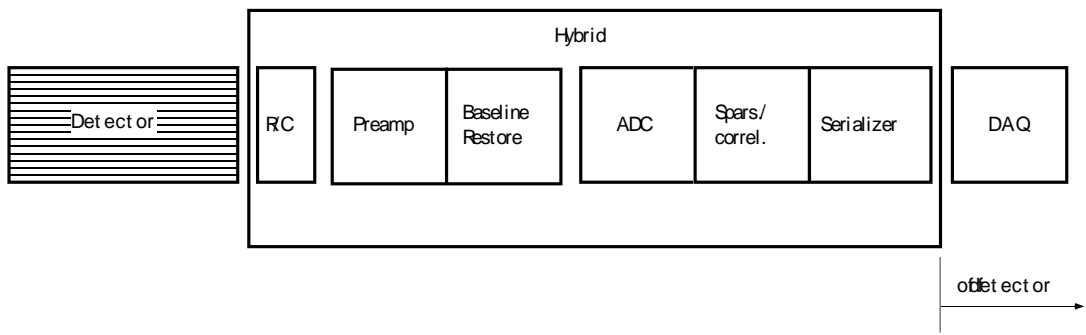,height=3.5in,%
bbllx=60bp,bblly=470bp,bburx=500bp,bbury=680bp,clip=}}
\vspace{-2cm}
\caption{\label{siel} A view of the silicon electronics design outline. The
bias resistors are off the detector.}
\end{figure}

The silicon electronic system is being developed along the same lines
as the RICH  electronics mentioned above. A possible system is outlined in
Fig.~\ref{siel}.

\section{Drift Chamber}

In order to reduce multiple scattering which limits the tracking resolution,
material in the chamber and the chamber walls is being reduced as much as
possible. To help photon detection, the endplate is being constructed
out of 1.5 cm thick shaped aluminum. To fit around the interaction region
quadrupoles, there is an
inner conical section which contains only axial wire layers and an outer
cylindrical section which contains alternating stereo wire layers
(see Fig.~\ref{CLEOIII}).

A Helium based gases will be used in the chamber. Properties of some
of these gases are shown in Table~\ref{table:dr} along with the ``standard",
50\% Ar-50\% CH$_4$ mixture. Here $R_L$ is the radiation length, $\theta_L$
the Lorentz angle, \#e/cm the number of ion pairs produced per cm,
$\sigma (dE/dx)$ the resolution in the specific ionization measurement and
$e/\pi$, the ratio of specific ionization between electrons and pions at
minimum $dE/dx$ loss. (The latter two quantities are from calculations.)

\begin{table}[th]
\centering
\caption{Comparison of Drift Chamber gas properties}
\label{table:dr}
\vspace*{2mm}
\begin{tabular}{rrrrrr}\hline\hline
%\multicolumn{2}{c}{$\BM$}&\multicolumn{2}{|c}{$\BP$}
Gas mixture & $R_L$ & $\theta_L$ & \#e/cm & $\sigma (dE/dx)$ &$e/\pi$ \\\hline
Ar-C$_2$H$_6$  (50:50) & 178 & 68.9 & 33.8 & 6.3\% & 1.46\\
He-C$_3$H$_8$  (60:40) & 569 & 35.0 & 32.6 & 6.3\% & 1.30 \\
He-C$_3$H$_8$  (40:60) & 392 & 38.6 & 45.8 & 5.7\% & 1.28 \\
He-C$_4$H$_(10)$  (70:30) &564 & 30.3 & 31.1 & 6.3\% & 1.29 \\
He-C$_2$H$_6$  (50:50) & 686 & 41.4 & 24.8 & 6.7\% & 1.32  \\
\end{tabular}
\end{table}

\begin{figure}[hbtp]
%\vspace{-2cm}
\centerline{\psfig{figure=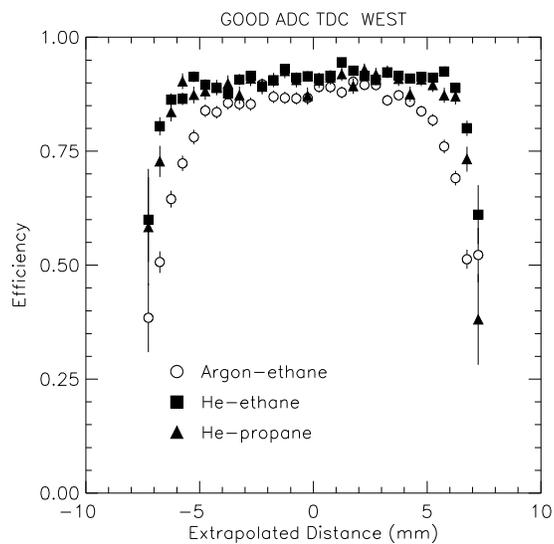,height=4in,%
bbllx=0bp,bblly=0bp,bburx=600bp,bbury=680bp,clip=}}
\vspace{-2cm}
\caption{\label{dr_eff} The hit efficiency across the drift cell for three
different gases. }
\end{figure}

We have found spatial
resolutions in test chambers of better than 100 $\mu$m for most of the He based
mixures, even when the test chambers have been placed in a 1.5T magnetic field.
These resolutions are better than achieved in Ar-C$_2$H$_6$ in the same
test setup. In Fig.~\ref{dr_eff} we show the hit efficiency for different
gases as function of the incident track position across the drift cell.
 The efficiency  is much better
using the He based gases due to the smaller Lorentz angle. These gases are
so promising that CLEO II  will immeadiately adopt one of the He mixtures.

\section{Acknowledgements}

This work reported here is supported by the National Science Foundation. I
thank M. Artuso,  C. Bebek,  H. Kagan and the RICH  group at Syracuse Univ. for
their help.


\begin{thebibliography} {99}
\bibitem{Pascos} S. Stone, ``Fundamental Constants from $b$ and $c$ Decay,"
               HEPSY 94-5,  in Proceedings of ``Particle Strings
               and Cosmology," meeting, Syracuse, NY (1994), ed. K. Wali,
               World Scientific, Singapore (1995) and in
                    ``Proceedings of DPF94 Meeting," Albuquerque, NM (1994) ed.
               S. Seidel, World Scientific, Singapore (1995).
\bibitem{pilnu} R. Ammar {\it et al.}, ``Measurement of the Branching Ratios
		for exclusive $b\to u\ell\nu$ Decays," CLEO CONF 95/9,
		EPS0165 (1995).
\bibitem{fDs} D. Gibaut {\it et al.}, ``Measurement of
		$\Gamma(D^+_s\to\mu^+\nu)/\Gamma(D^+_s\to\phi\pi^+)$,"
		CLEO  CONF 95-22, EPS0184 (1995).
\bibitem{rhogam} S. Playfer and S. Stone, ``Rare $B$ Decays", HEPSY 95-01
		(1995), to be published in Int. Journal of Mod. Phys. A.
\bibitem{CLEOII} Y. Kubota {\it et al.},  Nucl. Instr. and Meth. {\bf A301},
		{\bf A320} (1992) 56.
\bibitem{overv}
J. Seguinot and T. Ypsilantis, \nim{A343}{1}{1994}.
\bibitem{CERNt}
J.L. Guyonnet {\it et al.}, { Nucl. Instr. \& Meth.} {\bf A343}
(1994) 178.
\bibitem{alex}A. Efimov {\it et al.} , ``Monte Carlo Studies of a Novel LiF
Radiator for
RICH Detectors," HEPSY 94-8 (1994), to be published in Nucl. Instr. \& Meth.;
M. Artuso, ``The Ring Imaging Cherenkov Counter for
	CLEO III," presented at 6th Int. Symp. on Heavy Flavour Physics
		Pisa, Italy, June 1995, to appear in proceedings.

\bibitem{einar} {E. Nygard} {\it et al.},  Nucl. Instr. and Meth. {\bf A301}
(1991) 506.
\bibitem{chip} { O. Milgrome} {\it Talk given ath the 2nd International Meeting
 on Front End Electronics for Tracking Detectors at Future High Luminosity
  Colliders} Perugia, Italy (1994); R. Yarema, ``A Beginners Guide to the
  SVXII," FERMILAB-TM-1892 (1994).

\end{thebibliography}
\end{document}